\documentclass{article}
\usepackage{spconf,amsmath,graphicx,url}

\usepackage{booktabs}
\usepackage{color}
\usepackage{microtype}
\usepackage{xspace}
\usepackage{cite}
\usepackage{dingbat}
\usepackage{booktabs}
\usepackage{color}
\usepackage{balance}

\newcommand{\myparagraph}[1]{\medskip \noindent \textbf{#1}: }

\title{Speaker independence of neural vocoders and their effect on parametric resynthesis speech enhancement}
\twoauthors
 {Soumi Maiti}
    {The Graduate Center, CUNY\\
Computer Science\\
    New York, NY, USA \\
    smaiti@gradcenter.cuny.edu}
 {Michael I Mandel}
    {Brooklyn College \\
     Computer and Information Science\\
    Brooklyn, NY, USA \\
    mim@sci.brooklyn.cuny.edu}
\begin{document}
\ninept
\maketitle
\begin{abstract}
Traditional speech enhancement systems produce speech with compromised quality.  Here we propose to use the high quality speech generation capability of neural vocoders for better quality speech enhancement. We term this \emph{parametric resynthesis} (PR). In previous work, we showed that PR systems generate high quality speech for a single speaker using two neural vocoders, WaveNet and WaveGlow. Both these vocoders are traditionally speaker dependent. Here we first show that when trained on data from enough speakers, these vocoders can generate speech from unseen speakers, both male and female, with similar quality as seen speakers in training. 
Next using these two vocoders and a new vocoder LPCNet, we evaluate the noise reduction quality of PR on unseen speakers and show that objective signal and overall quality is higher than the state-of-the-art speech enhancement systems Wave-U-Net, Wavenet-denoise, and SEGAN.  Moreover, in subjective quality, multiple-speaker PR out-performs the oracle Wiener mask.
\end{abstract}
\section{Introduction}
\label{sec:intro}
Traditional speech enhancement systems modify a noisy mixture to reduce the amount of noise it contains, but in doing so they introduce distortion in the speech. The distortion increases when there is more noise in the mixture leading to poor quality speech~\cite{chen2006new}. In contrast, speech synthesis systems generate high quality speech from only textual information. These text-to-speech systems (TTS) are complex as they need to generate realistic acoustic representation \emph{without} a reference audio signal. 
In this work, we propose to combine these two methods, i.e., using  speech synthesis techniques for speech enhancement. This is an easier task than TTS since we have a reference noisy audio signal from which we can extract the desired prosody instead of having to invent it. By predicting the ``acoustic features'' of the clean speech from the noisy speech in the speech enhancement system, we can generate high quality noise-free resyntheses.

Parametric Resynthesis (PR) systems~\cite{maiti2019parametric, maiti2019speech} predict clean acoustic parameters from noisy speech and synthesize speech from these predicted parameters using a speech synthesizer or vocoder. Current speech synthesizers are trained to generate high quality speech for a single speaker. In previous work we showed that a single speaker PR system can synthesize very high quality clean speech at $22$~KHz~\cite{maiti2019parametric} and performs better than the corresponding TTS system~\cite{maiti2019speech}. Hence, a critical question is whether these systems can be generalized to unknown speakers. The main contribution of the current work is to show that when trained on a large number of speakers, neural vocoders can successfully generalize to unseen speakers. Furthermore, we show that PR systems using these neural vocoders can also generalize to unseen speakers in the presence of noise.

In this work, we test the speaker dependence of neural vocoders, and their effect on the enhancement quality of PR. We show that when trained on $56$ speakers, WaveGlow~\cite{prenger2018waveglow}, WaveNet~\cite{van2016wavenet}, and LPCNet~\cite{valin2019lpcnet} are able to generalize to unseen speakers.
We compare the noise reduction quality of PR with three state-of-the-art speech enhancement models and show that PR-LPCNet outperforms every other system including an oracle Wiener mask-based system. In terms of objective metrics, the proposed PR-WaveGlow performs better in objective signal and overall quality.

\subsection{Related work}

Traditional speech enhancement systems generally predict a time-Frequency mask to reduce noise in the magnitude spectrum domain, for example~\cite{WangTrainingTargetsSupervised2014, ErdoganPhasesensitiverecognitionboostedspeech2015}. Recent works perform speech enhancement in the time-domain directly, which has the additional advantage of reconstructing the phase of the signal. A modified WaveNet was proposed for speech denoising~\cite{rethage2018wavenet}, by using non-causal convolutions on noisy speech and predicting both clean speech and the noise signal. Another approach is to progressively downsample the noisy audio to a bottleneck feature and then upsample with skip connections to the corresponding downsampled features to enhance speech. SEGAN~\cite{pascual2017segan} uses this approach in a GAN setting and Wave-U-Net~\cite{macartney2018improved, stoller2018wave} uses it in the U-Net setting. The aim of these approaches is to remove noise from the audio at different scales. Compared to these systems, we do not focus on modelling noise but only focus on modelling speech.
We evaluate our approach against three of these systems~\cite{pascual2017segan,macartney2018improved,rethage2018wavenet}. These papers publish results on the same dataset we used and also each provide several enhanced files, which we utilize in our listening tests.

\section{System Overview}
\label{sec:tech_ovw}
Our PR models have two parts. First is a prediction model that estimates the clean acoustic features from noisy audio. Second, a vocoder synthesizes ``clean'' speech from the predicted ``clean'' acoustic parameters. The aim of the prediction model is to reduce noise while the vocoder synthesizes high quality audio.

\subsection{Prediction model}
\label{ssec:lstm_prediction_model}
The prediction model is trained with parallel clean and noisy speech. It takes noisy mel-spectrogram $Y$ as input and is trained to predict clean acoustic features $X$. The predicted clean acoustic features vary based on the vocoder used. In this work we used WaveGlow, WaveNet, LPCNet and WORLD~\cite{morise2016world} as vocoders. For Waveglow and WaveNet, we predict clean mel-spectrograms. For LPCNet, we predict $18$-dimensional Bark-scale frequency cepstral coefficients (BFCC) and two pitch parameters: period and correlation. For WORLD we predict the spectral envelope, aperiodicity, and pitch. For WORLD and LPCNet, we also predict the $\Delta$ and $\Delta\Delta$ of these acoustic features for smoother outputs.
The prediction model is trained to minimize the mean squared error (MSE) of the acoustic features
\begin{equation} \label{eq:pred_loss}
{\rm MSE:}\quad    \mathcal{L} = \| X - \hat{X}\|^2
\end{equation}
where $\hat{X}$ are the predicted and $X$ are the clean acoustic features. 
The Adam optimizer~\cite{KingmaAdamMethodStochastic2014} is used for training. During test, for a given a noisy mel-spectrogram, clean acoustic parameters are predicted. For LPCNet and WORLD we use maximum likelihood parameter generation (MLPG)~\cite{tokuda2000speech} algorithms to refine our estimate of the clean acoustic features from predicted acoustic features, $\Delta$, and $\Delta\Delta$.

\subsection{Vocoders}
\label{ssec:voc}
The second part of PR resynthesizes speech from the predicted acoustic parameters $\hat{X}$ using a vocoder. The vocoders are trained on clean speech samples $x$ and clean acoustic features $X$. During synthesis, we use predicted acoustic parameters $\hat{X}$ to generate predicted clean speech $\hat{x}$. 
In the rest of this section we describe the vocoders, three neural: WaveGlow, WaveNet, LPCNet and one non-neural: WORLD.
 
\myparagraph{WaveGlow}
\label{sssec:waveglow}
WaveGlow~\cite{prenger2018waveglow} is a Glow based network~\cite{kingma2018glow} for synthesizing speech. WaveGlow learns a sequence of invertible transformations of audio samples $x$ to a Gaussian distribution conditioned on the mel spectrogram $X$. For inference, WaveGlow samples a latent variable $z$ from the learned Gaussian distribution and applies the inverse transformations conditioned on $X$ to reconstruct the speech sample $\tilde{x}$. The model is trained to minimize the log likelihood of the clean speech
\begin{equation}\label{eq:waveglow_loss}
     \ln p(x \mid X) = \ln p(z) + \log\det \left| \frac{dz}{dx} \right|,
\end{equation} 
where $\ln p(z)$ is the log-likelihood of the spherical zero mean Gaussian with variance $\sigma^2$. During training $\sigma=1$ is used. We use the officially published waveGlow implementation\footnote{ \url{https://github.com/NVIDIA/waveglow}} with the original setup, i.e., 12 coupling layers, each consisting of 8 layers of dilated convolution with 512 residual and 256 skip connections. We will refer to the PR system with WaveGlow as its vocoder as PR-WaveGlow.

\myparagraph{LPCNet}
\label{sssec:lpcnet}
LPCNet is a variation of WaveRNN~\cite{kalchbrenner2018efficient} that simplifies the vocal tract response using linear prediction $p_t$ from previous time-step samples
\begin{equation*}
    p_t = \sum_{k=1}^{M} a_k x_{t-k}.
\end{equation*}
LPC coefficients $a_k$ are computed from the 18-band BFCC. It predicts the LPC predictor residual $e_t$, at time $t$. Then sample $x_t$ is generated by adding $e_t$ and $p_t$.

A frame conditioning feature $f$ is generated from $20$ input features: 18-band BFCC and 2 pitch parameters via two convolutional and two fully connected layers.  The probability $p(e_t)$ is predicted from $x_{t-1}$, $e_{t-1}$, $p_{t}$, $f$ via two GRUs~\cite{chung2014empirical} (A and B) combined with dualFC layer followed by a softmax. The largest GRU (GRU-A) weight matrix is forced to be sparse for faster synthesis. The model is trained on the categorical cross-entropy loss of $p(e_t)$ and the predicted probability of the excitation $\hat{p}(e_t)$.  Speech samples are $8$-bit mu-law quantized.
We use the officially published LPCNet implementation\footnote{ \url{https://github.com/mozilla/LPCNet}} with $640$ units in GRU-A and 16 units in GRU-B. We refer to the PR system with LPCNet as its vocoder as PR-LPCNet.

\myparagraph{WaveNet}
\label{sssec:wavenet}
WaveNet~\cite{van2016wavenet} is a autoregressive speech waveform generation model built with dilated causal convolutional layers. The generation of one speech sample at time step $t$, $x_t$, is conditioned on all previous time step samples $(x_1, x_2,\ldots x_{t-1})$. We use the Nvidia implementation\footnote{\url{https://github.com/NVIDIA/nv-wavenet}} which is the Deep-Voice~\cite{arik2017deep} model of WaveNet for faster synthesis. Speech samples are mu-law qauantized to 8 bits. The normalized log mel-spectrogram is used in local conditioning. 
WaveNet is trained on the cross-entropy between the quantized sample $x_{t}^{\mu}$ and the predicted quantized sample $\hat{x}_{t}^{\mu}$.

For WaveNet, we used a smaller model that is able to synthesize speech with moderate quality. We tested the PR model's dependency on speech synthesis quality by testing on a smaller model. We used $20$ layers with $64$ residual, $128$ skip connections, and 256 gate channels with maximum dilation of $128$. This model can synthesize clean speech with average predicted mean opinion score (MOS) $3.25$ for a single speaker~\cite{arik2017deep}.
The PR system with WaveNet as its vocoder is referred to as PR-WaveNet.

\myparagraph{WORLD}
\label{sssec:world}
Lastly, we use a non-neural vocoder WORLD which synthesizes speech from three acoustic parameters: spectral envelope, aperiodicity, and $F0$. We use WORLD with the Merlin toolkit \footnote{\url{https://github.com/CSTR-Edinburgh/merlin}}. WORLD is a source-filter model that takes previously mentioned parameters and synthesizes speech. We also use spectral enhancement to modify the predicted parameters as is standard in Merlin~\cite{wu2016merlin}.

\begin{table}[t]
    \centering
    \scriptsize
\begin{tabular}{l c| ccc| c }
\toprule
Model & \#spk & SIG & BAK & OVL & STOI \\
\midrule
\multicolumn{6}{l}{Seen}\\
\midrule
WaveGlow & 1 & 4.7$\pm$0.12 & 2.9$\pm$0.10 & 3.9$\pm$0.16  & 0.94$\pm$0.01 \\
LPCNet &  1 & 3.8$\pm$0.16 & 2.2$\pm$0.12 & 2.9$\pm$0.21 & 0.91$\pm$0.02 \\
 WaveNet & 1 &  3.3$\pm$0.15 & 2.1$\pm$0.06 & 2.5$\pm$0.13 & 0.81$\pm$0.03 \\
\midrule
\multicolumn{6}{l}{Unseen - Male}\\
\midrule
 WaveGlow & 3 & 4.4$\pm$0.03 & 2.8 $\pm$ 0.01& 3.7$\pm$0.02  & 0.94$\pm$0.01 \\
 LPCNet & 3 & 4.0$\pm$0.14  & 2.4$\pm$0.10 & 3.2$\pm$0.16  & 0.90$\pm$0.04\\
 WaveNet  & 3 & 3.2$\pm$0.08 & 2.1$\pm$0.07 & 2.5$\pm$0.10 & 0.83$\pm$0.01 \\
\midrule
\multicolumn{6}{l}{Unseen - Female}\\
\midrule
WaveGlow & 3 & 4.7$\pm$0.04 & 2.9$\pm$0.03 & 3.9$\pm$0.05  & 0.95$\pm$0.01 \\
LPCNet & 3 & 3.9$\pm$0.15  & 2.3$\pm$0.12 & 3.0$\pm$0.20  & 0.90$\pm$0.04\\
WaveNet  & 3 & 3.3$\pm$0.10 & 2.0$\pm$0.06 & 2.5$\pm$0.10 & 0.80$\pm$0.01 \\
\bottomrule
\end{tabular}
\caption{Speaker generalization of neural vocoders. Objective quality metrics for synthesis from true acoustic features, higher is better. Sorted by SIG. }
\label{tab:obj_clean}
\end{table}

\begin{table}[tb]
    \centering
    \footnotesize
\begin{tabular}{lccc c }
\toprule
Model & SIG & BAK & OVL & STOI \\
\midrule
Oracle Wiener & 4.3  & 3.8 & 3.9 &  0.98\\
\midrule
PR-WaveGlow &  3.8 & 2.4 & 3.1 & 0.91\\
PR-LPCNet, noisy $F0$ & 3.5 & 2.1 & 2.7 & 0.88 \\
PR-LPCNet  & 3.1  & 1.8 & 2.2 & 0.88 \\
PR-World  & 3.0 & 1.9 & 2.2 & 0.87 \\
PR-WaveNet   &  2.9 & 2.0 & 2.2 &  0.83\\
\midrule
Wave-U-Net (from~\cite{macartney2018improved}) & 3.5 & 3.2 & 3.0 & \_\\
SEGAN (from~\cite{pascual2017segan}) & 3.5 & 2.9 & 2.8 & \_\\
\bottomrule
\end{tabular}
\caption{Speech enhancement objective metrics on full 824-file test set:  higher is better.  Top system uses oracle clean speech information. Bottom section compares to published comparison system results.}
\label{tab:obj}
\end{table}

\section{Experiments}
\label{sec:experiments}
\subsection{Dataset}
We use the publicly available noisy VCTK dataset~\cite{valentini2017noisy} for our experiments. The dataset contains 56 speakers for training: 28 male and 28 female speakers from the US and Scotland. The test set contains two unseen voices, one male and another female. Further, there is another available training set, consisting 14 male and 14 female from England, which we use to test generalization to more speakers.


The noisy training set contains ten types of noise: two are artificially created, and the eight other are chosen from DEMAND~\cite{thiemann2013diverse}. The two artificially created are speech shaped noise and babble noise. The eight from DEMAND are noise from a kitchen, meeting room, car, metro, subway car, cafeteria, restaurant, and subway station. The noisy training files are available at four SNR levels: $15$, $10$, $5$, and $0$~dB. The noisy test set contains five other noises from DEMAND: living room, office, public square, open cafeteria, and bus. The test files have higher SNR: $17.5$, $12.5$, $7.5$, and $2.5$~dB. All files are downsampled to $16$~KHz for comparison with other systems. There are $23,075$ training audio files and $824$ testing audio files.

\begin{figure*}[t]
\begin{minipage}[b]{0.95\linewidth}
  \centering
  \centerline{\includegraphics[width=0.9\textwidth, trim={10cm, 0cm, 10cm, 1cm},clip]{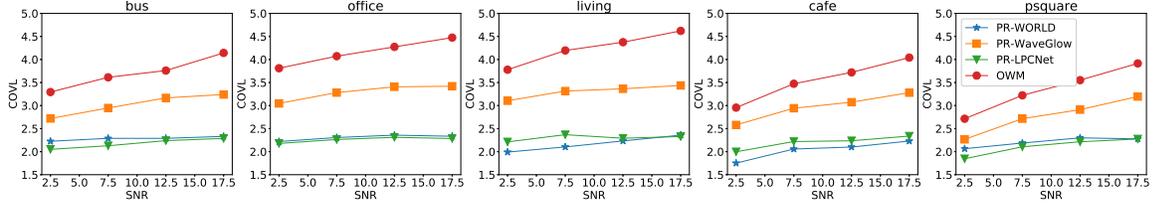}}
\end{minipage}
 \caption{ Overall objective quality of PR systems and OWM broken down by noise type (824 test files).}
\label{fig:noise_score}
\end{figure*}

\subsection{Exp 1: Speaker independence of neural vocoders}
Firstly, we test if WaveGlow and WaveNet can generalize to unseen speakers on clean speech. Using the data described above, we train both of these models with a large number of speakers ($56$) and test them on 6 unseen speakers. Next, we compare their performance to LPCNet which has previusly been shown to generalize to unseen speakers. In this test, each neural vocoder synthesizes speech from the original clean acoustic parameters. Following the three baseline papers~\cite{pascual2017segan, macartney2018improved, rethage2018wavenet}, we measure synthesis quality with objective enhancement quality metrics~\cite{hu2006evaluation} consisting of three composite scores: CSIG, CBAK, and COVL. These three measures are on a scale from 1 to 5, with higher being better. CSIG provides and estimate of the signal quality, BAK provides an estimate of the background noise reduction,  and OVL provides an estimate of the overall quality.

LPCNet is trained for 120 epochs with a batch size of 48, where each sequence has 15 frames. WaveGlow is trained for 500 epochs with batch size 4 utterances. WaveNet is trained for 200 epochs with batch size 4 utterances. For WaveNet and WaveGlow we use GPU synthesis, while for LPCNet CPU synthesis is used as it is faster\footnote{We also found that GPU synthesis code was incomplete as of commit \texttt{3a7ef33}}. WaveGlow and WaveNet synthesize from clean mel-spectrograms with window length 64~ms and hop size 16~ms. LPCNet acoustic features use a window size of 20~ms and a hop size of 10~ms.

We report the synthesis quality of three unseen male and three unseen female speakers, and compare them with unseen utterances from one known male speaker. For each speaker, the average quality is calculated over 10 files. Table~\ref{tab:obj_clean} shows the composite quality results along with the objective intelligibility score from STOI~\cite{taal2010short}. We observe that WaveGlow has the best quality scores in all the measures. The female speaker scores are close to the known speaker while the unseen male speaker scores are a little lower. We note here that these values are not as high as single speaker WaveGlow, which can synthesize speech very close to the ground truth. We also note that LPCNet scores are lower than those of WaveGlow but better than WaveNet. Between LPCNet and WaveNet, we do not observe a significant difference in synthesis quality for male and female voices. Although WaveNet has lower scores, it is consistent across known and unknown speakers. Thus, we can say that WaveNet generalizes to unseen speakers.

\begin{figure*}[t]
  \centering
  \centerline{\includegraphics[width=0.65\linewidth, trim={5.5cm 0.1cm
5.6cm 1.3cm},clip]{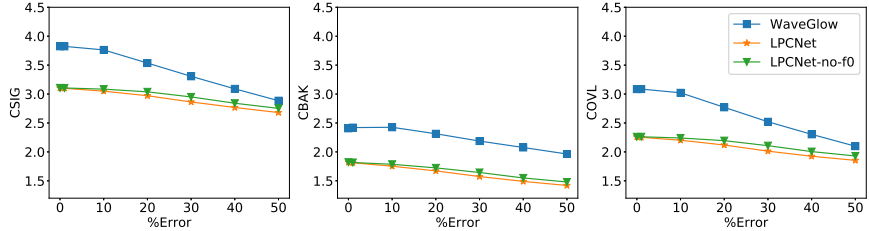}}
  \caption{Objective metrics as error is artificially added to the predictions of the acoustic features, higher is better. Error is measured as a proprotion of the standard deviation of the vocoders' acoustic features over time.}
  \label{fig:obj_error}
\end{figure*}

\begin{figure}[t]
  \centering
  \centerline{\includegraphics[width=0.8\linewidth, trim={0cm 0.6cm
0.8cm 0.65cm},clip]{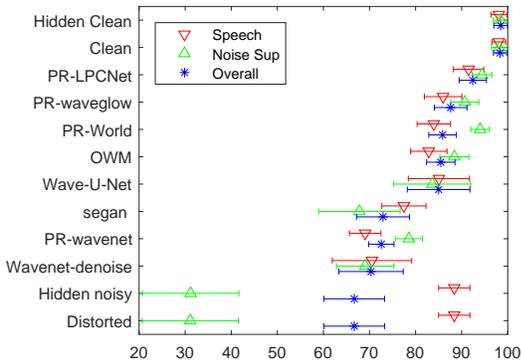}}
  \caption{Subjective quality: higher is better. Error bars show twice the standard error.}
  \label{fig:qual}
\end{figure}
\subsection{Exp 2: Speaker independence of parametric resynthesis}
Next, we test the generalizability of the PR system across different SNRs and unseen voices. We use the test set of 824 files with 4 different SNRs.
The prediction model is a 3-layer bi-directional LSTM with 800 units that is trained with a learning rate of $0.001$. For WORLD filter size is 1024 and hop length is 5~ms.
We compare PR models with a mask based oracle, the Oracle Wiener Mask (OWM), that has clean information available during test. 

Table~\ref{tab:obj} reports the objective enhancement quality metrics and STOI.  We observe that the OWM performs best, PR-WaveGlow performs better than Wave-U-Net and SEGAN on CSIG and COVL. PR-WaveGlow's CBAK score is lower, which is expected since this score is not very high even when we synthesize clean speech (as shown in Table~\ref{tab:obj_clean}). Among PR models, PR-WaveGlow scores best and PR-WaveNet performs worst in CSIG. The average synthesis quality of the WaveNet model affects the performance of the PR system poorly. PR-WORLD and PR-LPCNet scores are lower as well, we observe that both of these models sound much better than the objective scores would suggest. We believe, as both of these models predicts $F0$, even a slight error in $F0$ prediction affects the objective scores adversely. For this, we test the PR-LPCNet using the noisy $F0$ instead of the prediction, and the quality scores increase. In informal listening the subjective quality with noisy F0 is similar to or worse than the predicted $F0$ files. Hence we can say that the objective enhancement metrics are not a very good measure of quality for PR-LPCNet and PR-WORLD.

We also test objective quality of PR models and OWM against different SNR and noise types. The results are shown in Figure~\ref{fig:noise_score}. We observe with decreasing SNR, CBAK quality for PR models stays the same, while for OWM, CBAK score decreases rapidly. This shows that the noise has a smaller effect on background quality compared to a mask based system, i.e., the background quality is more related to the presence of synthesis artifacts than recorded background noise.  



\subsection{Listening tests}
\label{ssec:lis_test}
Next, we test the subjective quality of the PR systems with a listening test. For the listening test, we choose 12 of the 824 test files, with four files from each of the 2.5, 7.5 and 12.5~dB SNRs. We observed the 17.5~dB file to have very little noise, and all systems perform well with them. In the listening test, we also compare with the OWM and three comparison models. For these comparison systems, we included the publicly available output files in our listening tests, selecting five files from each: Wave-U-Net has 3 from 12.5~dB and 2 from 2.5~dB, Wavenet-denoise and SEGAN have 2 common files from 2.5~dB, 2 more files each are selected from 7.5 dB and 1 from 12.5 dB. For Wave-U-Net, there were no 7.5 dB files available publicly.

The listening test follows the Multiple Stimuli with Hidden Reference and Anchor (MUSHRA) paradigm~\cite{MUSHRA}. Subjects were presented with 8-10 anonymized and randomized versions of each file to facilitate direct comparison: 4 PR systems (PR-WaveNet, PR-WaveGlow, PR-LPCNet, PR-World), 4 comparison speech enhancement systems (OWM, Wave-U-Net, WaveNet-denoise, and SEGAN), and clean and noisy signals. Subjects were also provided reference clean and noisy versions of each file\footnote{All files are available at   \url{http://mr-pc.org/work/icassp20/}}. Five subjects took part in the listening test. They were told to rate the speech quality, noise-suppression quality, and overall quality of the speech from $0-100$, with $100$ being the best. 
We observe intelligibility of all of the files to be very high, so instead of doing an intelligibility listening test, we ask subjects to rate the subjective intelligibility as a score from $0-100$.

\begin{table}[bt]
    \centering
    \footnotesize
\begin{tabular}{lccc c c}
\toprule
Model & SIG & BAK & OVL & STOI & Subj.~Intel. \\
\midrule
Oracle Wiener & 4.3  & 3.8 & 3.9 &  0.98 & 0.91 \\
\midrule
PR-WaveGlow &  3.7 & 2.4 & 3.0 & 0.91 & 0.90 \\
PR-World  & 3.0 & 1.9 & 2.2 & 0.86 & 0.90\\
PR-LPCNet  & 3.0  & 1.8 & 2.2 & 0.85 & 0.92\\
PR-WaveNet   &  2.9 & 2.0 & 2.2 &  0.83 & 0.74\\
\bottomrule
\end{tabular}
\caption{Speech enhancement objective metrics and subjective intelligibility on the 12 listening test files.}
\label{tab:sub_lis_test}
\end{table}

Figure~\ref{fig:qual} shows the result of the quality listening test. PR-LPCNet performs best in all three quality scores, followed by PR-WaveGlow and PR-World.  The next best model is the Oracle Wiener mask followed by Wave-U-Net. 
Table~\ref{tab:sub_lis_test} shows the subjective intelligibility ratings, where PR-LPCNet has the highest subjective intelligibility, followed by OWM, PR-WaveGlow, and PR-World. 
It also reports the objective quality metrics on the 12 files selected for the listening test for comparison with Table~\ref{tab:obj} on the full test set. We observe that while PR-LPCNet and PR-WORLD have very similar objective metrics (both quality and intelligibility), they have very different subjective metrics, with PR-LPCNet being rated much higher).

\subsection{Tolerance to error}
Finally, we measure the tolerance of PR models to inaccuracy of the prediction LSTM using the two best performing vocoders, WaveGlow and LPCNet. For this test, we randomly select 30 noisy test files. 
We make the predicted feature $\hat{X}$ noisy as, $\hat{X}_{e} = \hat{X}+\epsilon N$, where $\epsilon=MSE \times e\%$. The random noise $N$ is generated from a Gaussian distribution with the same mean and variance at each freuency as $X$.  Next, we synthesize with the vocoder from $\hat{X}_{e}$. For WaveGlow, $X$ is the mel-spectrogram and for LPCNet, $X$ is 20 features. We repeat the LPCNet test adding noise into all features and only the 18 BFCC features (not adding noise to $F0$). 

Figure~\ref{fig:obj_error} shows the objective metrics for these files. We observe that for WaveGlow, $e = 0-10\%$ does not affect the synthesis quality very much and $e>10\%$ decreases performance incrementally. For LPCNet, we observe that errors in the BFCC are tolerated better than errors in $F0$.

\section{Conclusion}
\label{sec:conclusion}
We show that the neural vocoders WaveGlow, WaveNet, and LPCNet can be used for speaker-independent speech synthesis when trained on 56 speakers. We also show that using these three vocoders, the parametric resynthesis model is able to generalize to new noises and new speakers across different SNRs. We find that PR-LPCNet outperforms the oracle Wiener mask-based system in subjective quality.

\bibliographystyle{IEEEbib}
\bibliography{strings_long,refs}

\begin{thebibliography}{10}

\bibitem{chen2006new}
Jingdong Chen, Jacob Benesty, Yiteng Huang, and Simon Doclo,
\newblock ``New insights into the noise reduction wiener filter,''
\newblock {\em {IEEE} Transactions on Audio, Speech, and Language Processing},
  vol. 14, no. 4, pp. 1218--1234, 2006.

\bibitem{maiti2019parametric}
Soumi Maiti and Michael~I Mandel,
\newblock ``Parametric resynthesis with neural vocoders,''
\newblock in {\em {IEEE} Workshop on Applications of Signal Processing to Audio
  and Acoustics}, 2019,
\newblock To appear.

\bibitem{maiti2019speech}
Soumi Maiti and Michael~I Mandel,
\newblock ``Speech denoising by parametric resynthesis,''
\newblock in {\em Proceedings of the {IEEE} International Conference on
  Acoustics, Speech, and Signal Processing}. IEEE, 2019, pp. 6995--6999.

\bibitem{prenger2018waveglow}
Ryan Prenger, Rafael Valle, and Bryan Catanzaro,
\newblock ``Waveglow: A flow-based generative network for speech synthesis,''
\newblock {\em arXiv preprint arXiv:1811.00002}, 2018.

\bibitem{van2016wavenet}
A{\"a}ron van~den Oord, Sander Dieleman, Heiga Zen, Karen Simonyan, Oriol
  Vinyals, Alex Graves, Nal Kalchbrenner, Andrew~W Senior, and Koray
  Kavukcuoglu,
\newblock ``{WaveNet}: A generative model for raw audio.,''
\newblock in {\em Proc.~ISCA SSW}, Sept. 2016, p. 125.

\bibitem{valin2019lpcnet}
Jean-Marc Valin and Jan Skoglund,
\newblock ``Lpcnet: Improving neural speech synthesis through linear
  prediction,''
\newblock in {\em Proceedings of the {IEEE} International Conference on
  Acoustics, Speech, and Signal Processing}. IEEE, 2019, pp. 5891--5895.

\bibitem{WangTrainingTargetsSupervised2014}
Yuxuan Wang, Arun Narayanan, and DeLiang Wang,
\newblock ``On training targets for supervised speech separation,''
\newblock {\em {IEEE} Transactions on Audio, Speech, and Language Processing},
  vol. 22, no. 12, pp. 1849--1858, 2014.

\bibitem{ErdoganPhasesensitiverecognitionboostedspeech2015}
Hakan Erdogan, John~R. Hershey, Shinji Watanabe, and Jonathan Le~Roux,
\newblock ``Phase-sensitive and recognition-boosted speech separation using
  deep recurrent neural networks,''
\newblock in {\em Proceedings of the {IEEE} International Conference on
  Acoustics, Speech, and Signal Processing}, 2015, vol. 2015-Augus.

\bibitem{rethage2018wavenet}
Dario Rethage, Jordi Pons, and Xavier Serra,
\newblock ``A wavenet for speech denoising,''
\newblock in {\em Proceedings of the {IEEE} International Conference on
  Acoustics, Speech, and Signal Processing}, 2018, pp. 5069--5073.

\bibitem{pascual2017segan}
Santiago Pascual, Antonio Bonafonte, and Joan Serr{\`a},
\newblock ``Segan: Speech enhancement generative adversarial network,''
\newblock {\em arXiv preprint arXiv:1703.09452}, 2017.

\bibitem{macartney2018improved}
Craig Macartney and Tillman Weyde,
\newblock ``Improved speech enhancement with the wave-u-net,''
\newblock {\em arXiv preprint arXiv:1811.11307}, 2018.

\bibitem{stoller2018wave}
Daniel Stoller, Sebastian Ewert, and Simon Dixon,
\newblock ``Wave-u-net: A multi-scale neural network for end-to-end audio
  source separation,''
\newblock {\em arXiv preprint arXiv:1806.03185}, 2018.

\bibitem{morise2016world}
Masanori Morise, Fumiya Yokomori, and Kenji Ozawa,
\newblock ``{WORLD}: a vocoder-based high-quality speech synthesis system for
  real-time applications,''
\newblock {\em IEICE Transactions on Information and Systems}, vol. 99, no. 7,
  pp. 1877--1884, Jul. 2016.

\bibitem{KingmaAdamMethodStochastic2014}
Diederik~P. Kingma and Jimmy Ba,
\newblock ``Adam: {{A Method}} for {{Stochastic Optimization}},''
\newblock {\em arXiv:1412.6980 [cs]}, Dec. 2014.

\bibitem{tokuda2000speech}
Keiichi Tokuda, Takayoshi Yoshimura, Takashi Masuko, Takao Kobayashi, and
  Tadashi Kitamura,
\newblock ``Speech parameter generation algorithms for hmm-based speech
  synthesis,''
\newblock in {\em Proceedings of the {IEEE} International Conference on
  Acoustics, Speech, and Signal Processing}. IEEE, 2000, vol.~3, pp.
  1315--1318.

\bibitem{kingma2018glow}
Diederik~P Kingma and Prafulla Dhariwal,
\newblock ``Glow: Generative flow with invertible 1x1 convolutions,''
\newblock {\em arXiv preprint arXiv:1807.03039}, 2018.

\bibitem{kalchbrenner2018efficient}
Nal Kalchbrenner, Erich Elsen, Karen Simonyan, Seb Noury, Norman Casagrande,
  Edward Lockhart, Florian Stimberg, Aaron van~den Oord, Sander Dieleman, and
  Koray Kavukcuoglu,
\newblock ``Efficient neural audio synthesis,''
\newblock {\em arXiv preprint arXiv:1802.08435}, 2018.

\bibitem{chung2014empirical}
Junyoung Chung, Caglar Gulcehre, KyungHyun Cho, and Yoshua Bengio,
\newblock ``Empirical evaluation of gated recurrent neural networks on sequence
  modeling,''
\newblock {\em arXiv preprint arXiv:1412.3555}, 2014.

\bibitem{arik2017deep}
Sercan~{\"O} Arik, Mike Chrzanowski, Adam Coates, Gregory Diamos, Andrew
  Gibiansky, Yongguo Kang, Xian Li, John Miller, Andrew Ng, Jonathan Raiman,
  et~al.,
\newblock ``Deep voice: Real-time neural text-to-speech,''
\newblock in {\em Proceedings of the International Conference on Machine
  Learning}. JMLR. org, 2017, pp. 195--204.

\bibitem{wu2016merlin}
Zhizheng Wu, Oliver Watts, and Simon King,
\newblock ``Merlin: An open source neural network speech synthesis system,''
\newblock {\em Proc. SSW}, 2016.

\bibitem{valentini2017noisy}
Cassia Valentini-Botinhao et~al.,
\newblock ``Noisy speech database for training speech enhancement algorithms
  and tts models,''
\newblock {\em University of Edinburgh. School of Informatics. Centre for
  Speech Technology Research (CSTR)}, 2017.

\bibitem{thiemann2013diverse}
Joachim Thiemann, Nobutaka Ito, and Emmanuel Vincent,
\newblock ``The diverse environments multi-channel acoustic noise database
  (demand): A database of multichannel environmental noise recordings,''
\newblock in {\em Proceedings of Meetings on Acoustics ICA2013}. ASA, 2013,
  vol.~19, p. 035081.

\bibitem{hu2006evaluation}
Yi~Hu and Philipos~C Loizou,
\newblock ``Evaluation of objective measures for speech enhancement,''
\newblock in {\em Proceedings of Interspeech}, 2006.

\bibitem{taal2010short}
Cees~H Taal, Richard~C Hendriks, Richard Heusdens, and Jesper Jensen,
\newblock ``A short-time objective intelligibility measure for time-frequency
  weighted noisy speech,''
\newblock in {\em Proceedings of the {IEEE} International Conference on
  Acoustics, Speech, and Signal Processing}, 2010, pp. 4214--4217.

\bibitem{MUSHRA}
``Method for the subjective assessment of intermediate quality level of audio
  systems,''
\newblock Tech. {R}ep. BS.1534-3, International Telecommunication Union
  Radiocommunication Standardization Sector ({ITU-R}), 2015.

\end{thebibliography}

\end{document}